\documentclass[showpacs,amsmath,
reprint,
aps,prl]{revtex4-1}

\usepackage{graphicx}
\usepackage{amsmath}
\begin{document}

\title
{Spontaneous exciton dissociation in carbon nanotubes}
\author{Y.~Kumamoto}
\author{M.~Yoshida}
\author{A.~Ishii}
\author{A.~Yokoyama}
\author{T.~Shimada}
\author{Y.~K.~Kato}
\email[Corresponding author. ]{ykato@sogo.t.u-tokyo.ac.jp}
\affiliation{Institute of Engineering Innovation, 
The University of Tokyo, Tokyo 113-8656, Japan}

\begin{abstract}
Simultaneous photoluminescence and photocurrent measurements on individual single-walled carbon nanotubes reveal spontaneous dissociation of excitons into free electron-hole pairs. Correlation of luminescence intensity and photocurrent shows that a significant fraction of excitons are dissociating during their relaxation into the lowest exciton state. Furthermore, the combination of optical and electrical signals also allows for extraction of the absorption cross section and the oscillator strength. Our observations explain the reasons for photoconductivity measurements  in single-walled carbon nanotubes being straightforward despite the large exciton binding energies.
\end{abstract}

\maketitle

Enhancement of the Coulomb interaction occurs in one dimensional systems because of limited screening \cite{Ogawa:1991}, and single-walled carbon nanotubes (SWCNTs) are an ideal model system where such an effect manifests itself \cite{Ando:1997}. Electron-hole pairs form tightly-bound excitons with a binding energy of a few hundred meV, which amounts to a significant fraction of the band gap energy \cite{Wang:2005, Maultzsch:2005}. Such a large binding energy warrants the stability of excitons even at room temperature, and with exciton size being a few nm \cite{Luer:2009, Schoppler:2011}, strong fields on the order of 100~V/$\mu$m would be required for exciton dissociation \cite{Perebeinos:2007}. 

In contrast to the expectation that generation of free carriers from charge-neutral excitons would be difficult, photocurrent and photovoltaic measurements have proved to be simple and convenient tools for studying the properties of SWCNTs. Not only have they been used to measure potential landscapes \cite{Balasubramanian:2005, Freitag:2007apl, Ahn:2007, Rauhut:2012}, optical absorption properties \cite{Lee:2007apl, Mohite:2008, Barkelid:2012}, and ultrafast carrier dynamics \cite{Prechtel:2011}, they have been instrumental in investigating unique effects that occur in SWCNTs, such as band-gap renormalization \cite{Lee:2007prb} and multiple electron-hole pair generation \cite{Gabor:2009}. It has been a perplexing situation where exciton dissociation has not been brought up as an obstacle for performing these experiments. In interpreting the results, quantitative discussion on the dissociation process has been scarce, and in some cases the excitonic effects have not been considered at all. 

Here we resolve such an inconsistency by performing simultaneous photoluminescence (PL) and photocurrent (PC) measurements on individual SWCNTs. Non-zero photoconductivity is observed even at small fields, indicating that excitons are spontaneously dissociating. A simple model is constructed to consistently describe the excitation power and voltage dependences of the PL and PC. Using this model, we find that a good fraction, if not majority, of excitons are dissociating into free carriers. Within the same analysis framework, we are also able to extract the absorption cross section and the oscillator strength at the $E_{22}$ resonance.

\begin{figure}[b]
\includegraphics{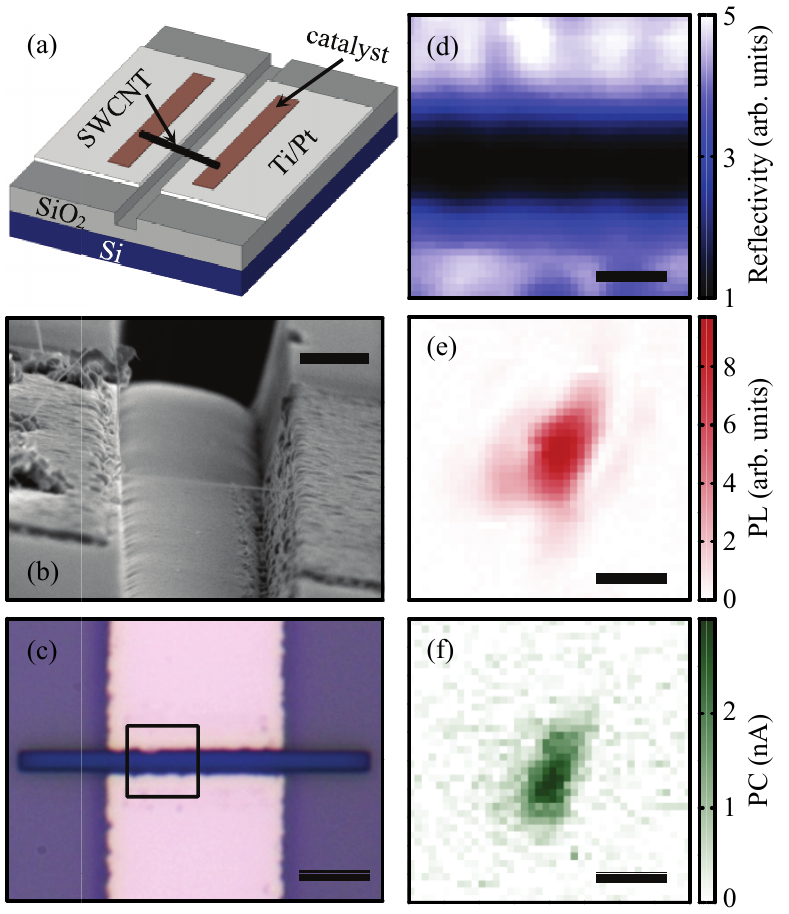}
\caption{\label{fig1}
(a) A schematic of a device.
(b) A scanning electron micrograph of a typical device.
(c) An optical microscope image of a device with a trench width of 1.3~$\mu$m. The black box shows the scan area for imaging measurements shown in (d)-(f).
The scale bars in (b) and (c) are 0.5~$\mu$m and 4~$\mu$m, respectively.
(d), (e), and (f) are reflectivity, PL, and PC images, respectively. The scale bars are 1~$\mu$m. Excitation energy and bias voltage are 1.651~eV and 20~V, respectively, and laser polarization angle is adjusted to maximize the PL signal. 
For (e), the PL image is extracted at an emission energy of 922~meV with a spectral integration window of 7~meV.
}\end{figure}

Our devices are field effect transistors with individual air-suspended SWCNTs \cite{Yasukochi:2011} as shown in Fig.~\ref{fig1}(a). We start with a Si substrate with 1-$\mu$m-thick oxide, and etch $\sim$500-nm-deep trenches into the oxide layer. An electron beam evaporator is used to deposit 3-nm Ti and 45-nm Pt for electrodes. Finally, catalyst particles are placed on the contacts and alcohol chemical vapor deposition is performed to grow SWCNTs \cite{Maruyama:2002, Imamura:2013}. A scanning electron micrograph of a typical device is shown in Fig.~\ref{fig1}(b). 

We look for devices that show nanotube PL at the trench in between the electrodes using a confocal microscope \cite{Moritsubo:2010, Watahiki:2012}. A continuous-wave Ti:sapphire laser is used for excitation and PL is detected by an InGaAs photodiode array attached to a spectrometer. The PC measurements are performed by monitoring the current through the device in the presence of a bias voltage $V$. We apply $-V/2$ and $+V/2$ to the two contacts, respectively, and ground the Si substrate. Although we do not expect much electrostatic doping because of the relatively thick oxide, this configuration ensures that the effective gate voltage at the center of the trench is zero. The current is averaged while a PL spectrum is collected, and the PC is obtained by subtracting the dark current measured in a similar manner with the laser blocked by a shutter. All measurements are done in air at room temperature.

Figure~\ref{fig1}(c) is an optical microscope image of the device, and in the area indicated by the black box,  we perform reflectivity, PL, and PC imaging simultaneously at an excitation laser power $P=15$~$\mu$W. The reflectivity image [Fig.~\ref{fig1}(d)] shows the position of the trench, and a luminescent nanotube suspended over the trench can be seen in the PL image [Fig.~\ref{fig1}(e)]. The PC image shows that the signal is maximized at the same spot as PL [Fig.~\ref{fig1}(f)]. In contrast to the case of Schottky barrier imaging \cite{Balasubramanian:2005, Freitag:2007apl, Ahn:2007, Rauhut:2012}, we do not observe PC when the laser spot is near or on the contacts. This confirms that band bending and electrostatic doping near the contacts are negligible in our voltage configuration.

\begin{figure}
\includegraphics{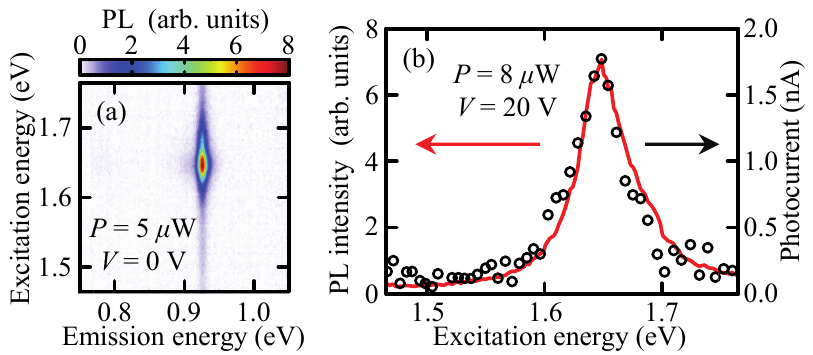}
\caption{\label{fig2}
(a) A PL excitation map for the same nanotube as shown in Fig.~\ref{fig1}(d-f) for $P=5$~$\mu$W and $V=0$~V. 
(b) PL (red curve) and PC (open circles) spectra taken with $P=8$~$\mu$W and $V=20$~V. Laser polarization is parallel to the nanotube axis. PL intensity is obtained by fitting the emission spectra with Lorentzian functions and taking the peak area. 
}\end{figure}

PL excitation spectroscopy performed on this nanotube at zero bias voltage shows a clear single peak [Fig.~\ref{fig2}(a)], and we identify the nanotube chirality to be $(10,6)$. By performing such an excitation spectroscopy under an application of bias, we obtain PL and PC excitation spectra simultaneously [Fig.~\ref{fig2}(b)]. Both PL and PC have a peak at the same excitation energy corresponding to the $E_{22}$ resonance. The spatial and spectral coincidence of the PL and PC signals show that both are indeed coming from the same nanotube.

\begin{figure*}
\includegraphics{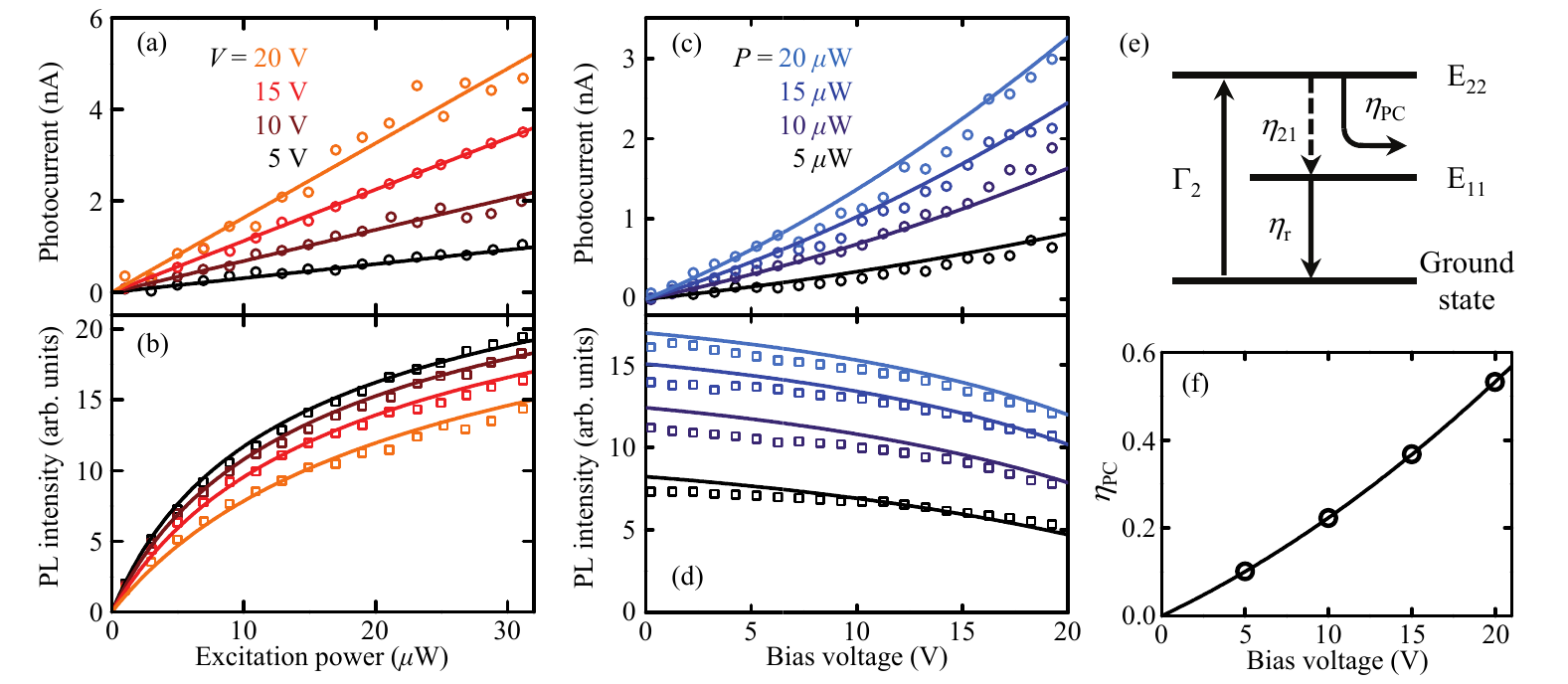}
\caption{\label{fig3}
(a) Excitation power dependence of PC. Data from bottom to top correspond to $V=5$, 10, 15, and 20~V.
(b) Power dependence of PL, with data from top to bottom corresponding to $V=5$, 10, 15, and 20~V.
(c) and (d) Bias voltage dependence of PC and PL, respectively. Data from  bottom to top correspond to $P=5$, 10, 15, and 20~$\mu$W.
For (a-d), the same tube as shown in Fig.~\ref{fig1}(d-f) was measured with the laser spot at the center of the nanotube. The excitation energy is fixed at 1.651~eV  and the laser polarization is parallel to the nanotube axis. Symbols are data and lines are simulation results as explained in the text.
(e) A schematic of the model used to produce the curves shown in (a-d).
(f) $\eta_\text{PC}$ as a function of $V$. Open circles are data obtained from (b) and the line is a fit as explained in the text.}\end{figure*}

On this device, the excitation power and bias voltage dependences are investigated in Fig.~\ref{fig3}(a-d). We first discuss the excitation power dependence. For all of the voltages, the PC signal shows a linear increase with excitation power [Fig.~\ref{fig3}(a)], whereas PL shows a sublinear increase [Fig.~\ref{fig3}(b)]. The latter behavior is known to be caused by exciton-exciton annihilation \cite{Matsuda:2008, Murakami:2009b, Xiao:2010, Moritsubo:2010}. If the observed PC is caused by dissociation of the $E_{11}$ excitons, then we expect PC to scale with PL, as both of the signals should be proportional to the number of $E_{11}$ excitons.

Rather, the linear behavior suggests that the PC is proportional to the number of excitons injected at the $E_{22}$ energy, and that dissociation of $E_{11}$ excitons is negligible. There are at least two different processes that can result in the dissociation of $E_{22}$ excitons. It is possible that the applied electric field induces the dissociation, and in this case one would expect some threshold voltage at which the dissociation occurs \cite{Perebeinos:2007}. Another conceivable scenario is the dissociation that happens spontaneously in the course of relaxation down to $E_{11}$ exciton states. 

The two pictures can be distinguished by examining the voltage dependence of the PC [Fig.~\ref{fig3}(c)]. We observe that the PC has a slightly superlinear dependence on the applied voltage, but there exists some slope near $V=0$. This implies that the conductivity is non-zero even at zero applied bias, supporting the interpretation that the injected excitons are spontaneously dissociating.

We note that the lack of field-induced dissociation for $E_{22}$ excitons is consistent with the interpretation of the intensity dependences that $E_{11}$ exciton dissociation is negligible. The binding energy for $E_{22}$ excitons is larger than $E_{11}$ excitons \cite{Perebeinos:2004}, and therefore we do not expect field-induced dissociation of $E_{22}$ excitons if $E_{11}$ excitons are still intact.

The voltage dependence of the PL [Fig.~\ref{fig3}(d)] shows decrease of PL with increasing voltage. Because of more photocarriers that flow into the contacts at higher voltages, we do expect that less excitons relax into the $E_{11}$ states. As the current gives the absolute rate of electron-hole pairs extracted from the nanotube, we can deduce the number of excitons removed from the system. By modeling such a fractioning in the exciton population, we are able to determine the number of injected excitons, and in turn the absorption cross section.

Figure~\ref{fig3}(e) shows a schematic of our model. $E_{22}$ excitons are generated at a rate 
\begin{equation}
\Gamma_2=\int n \sigma \frac{2P}{\pi r^2 E} \exp(-2 \frac{x^2}{r^2}) dx = \sqrt{\frac{2}{\pi}} \frac{n}{rE}\sigma P,
\end{equation}
where $n=130$~nm$^{-1}$ is the number of atoms per length, $\sigma$ is the absorption cross section per carbon atom, $r=492$~nm is the $1/e^2$ radius of the laser spot, and $E$ is the laser photon energy. The fraction of the excitons that are extracted by PC is denoted by $\eta_\text{PC}$, while $\eta_{21}=1-\eta_\text{PC}$ represents the fraction that relax down to the $E_{11}$ sublevel. The fraction of the $E_{11}$ excitons that recombine radiatively and contribute to PL is represented by a non-linear function $\eta_r(\Gamma_1)$ which includes the effects of exciton-exciton annihilation. Here, $\Gamma_1=\Gamma_2 \eta_{21}$ is the rate at which the $E_{11}$ excitons are populated.

The absolute values of $\eta_{21}$ can be obtained from the excitation power dependence of PL [Fig.~\ref{fig3}(b)]. At $V=0$, there are no PC and therefore $\eta_\text{PC}=0$ and $\eta_{21}=1$. When voltages are applied, $\Gamma_1$ decreases by a factor $\eta_{21}$. By scaling the excitation power to match the dependence at $V=0$, the values of $\eta_{21}$ are obtained for the four voltages. We plot $\eta_\text{PC}=1-\eta_\text{21}$ in Fig.~\ref{fig3}(f).

Having obtained the explicit values of $\eta_\text{PC}$, we can now determine $\sigma$. Within our model, the PC is given by 
\begin{equation}
I=e \eta_\text{PC} \Gamma_2=\sqrt{\frac{2}{\pi}} \frac{e \eta_\text{PC} n}{rE} \sigma P,
\end{equation}
where $e$ is the electron charge, and the only unknown parameter is $\sigma$. We find that a value of $\sigma=2.4\times 10^{-17}$~cm$^2$ best matches the PC data in Fig.~\ref{fig3}(a). This value is comparable to recent measurements of $\sigma$ at the $E_{22}$ resonance in micelle-encapsulated tubes \cite{Oudjedi:2013} and on-substrate tubes \cite{Joh:2011}. 

In addition to $\sigma$, the oscillator strength $f$ is obtained using its relation to integrated absorption cross section \cite{Sakurai}. We fit the $E_{22}$ resonance with a Lorentzian profile and obtain a linewidth of $\hbar \gamma=44.5$~meV where $\hbar$ is the Planck constant, and we use $f=\epsilon_0 m c \sigma \gamma/e^2$, where $\epsilon_0$ is the vacuum permittivity, $m$ is the electron mass, and $c$ is the speed of light. We find $f=0.015$ which is somewhat larger compared to (6,5) nanotubes \cite{Schoppler:2011}.

To verify the validity of our model, we simulate the intensity and voltage dependences of PC and PL using the parameters obtained above. For the voltage dependence of $\eta_\text{PC}$, we fit the data in Fig.~\ref{fig3}(f) with a linear term and a quadratic term. We use an analytic expression derived in Ref.~\cite{Murakami:2009b} for the form of $\eta_r(\Gamma_1)$, with the parameters adjusted to fit our data. As shown as solid lines in Fig.~\ref{fig3}(a-d), the model consistently explains all the data simultaneously.

The behavior of $\eta_\text{PC}$ shows that a large fraction of the injected excitons are dissociating, reaching a value as high as $\eta_\text{PC}=0.53$ at $V=20$~V. We expect PC to saturate above a certain voltage when all free carriers are extracted, but we do not see any signs of such saturation. This suggests that there are much more free carriers available even at the highest bias voltage we used, implying that the majority of the injected excitons are dissociating.

In order to check the reproducibility and to obtain $\sigma$ for other chiralities, we have performed similar measurements on other devices and the results are summarized in Table~\ref{tab1}. For four tubes with a chirality of $(8,7)$, we find that $f$ falls within $\pm 20\%$. We have observed that $\sigma$ can differ by a factor of three or so for other chiralities. 

\begin{table}
\caption{\label{tab1}
Absorption cross section and oscillator strength for the eight nanotubes measured. Lorentzian fits to PL excitation spectra at $V=0$~V are used to obtain the $E_{22}$ energy and full-width at half-maximum $\hbar \gamma$.
}
\begin{ruledtabular}
\begin{tabular}{cccccc}
Chirality & $E_{22}$ & $\hbar \gamma$ &$\sigma$ & $f$ \\
 &  (eV)  &  (meV) & ($\times 10^{-17}$~cm$^2$) & \\ \hline
(8,7) & 1.724  & 66.6  & 2.1  & 0.020\\
(8,7) & 1.712  & 58.4  & 2.6  & 0.022\\
(8,7) & 1.717  & 71.3  & 1.7  & 0.017\\
(8,7) & 1.725  & 69.1  & 2.5  & 0.025\\
(9,7) & 1.593  & 44.2  & 9.5  & 0.060\\
(9,8) & 1.555  & 50.5  & 7.1  & 0.052\\
(10,6) & 1.652 & 44.5   & 2.4  & 0.015\\
(10,8) & 1.452 & 51.5  & 1.3  & 0.009\\
\end{tabular}
\end{ruledtabular}
\end{table}

We note that our model does not consider any direct recombination of $E_{22}$ excitons which occurs prior to relaxation to the $E_{11}$ state, for example exciton-exciton annihilation at the $E_{22}$ level \cite{Harrah:2011b}. Such a process would lead to an underestimate of the number of injected excitons, and $\sigma$ would be larger than what we have deduced from our model. In addition, free carrier generation from $E_{11}$ exciton-exciton annihilation \cite{Santos:2011} is not taken into account explicitly. In principle, such a process can be identified by the behavior of PC at low powers, but the strong exciton-exciton annihilation in air-suspended nanotubes \cite{Xiao:2010, Moritsubo:2010} makes such an identification difficult. Further measurements at different excitation energies are expected to illuminate the relaxation kinetics of excitons.

In summary, we have performed simultaneous PL and PC spectroscopy on individual SWCNTs and constructed a model that consistently explains the excitation power and voltage dependences. Within the voltage range explored, we did not find evidences of field-induced exciton dissociation, for either of the $E_{11}$ and $E_{22}$ excitons. Instead, a considerable fraction of the injected excitons are found to spontaneously dissociate into free electron-hole pairs.  We have also obtained the absorption cross section and the oscillator strength from these air-suspended SWCNTs. Our findings explain why the large exciton binding energies do not impede photoconductivity measurements in SWCNTs.

\begin{acknowledgments}
We thank R. Saito, Y. Miyauchi and S. Maruyama for helpful discussions, T. Kan and I. Shimoyama for the use of the evaporator, and S. Chiashi and S. Maruyama for the use of the electron microscope. Work supported by KAKENHI (21684016, 23104704, 24340066, 24654084), SCOPE, and KDDI Foundation, as well as the Photon Frontier Network Program of MEXT, Japan. The devices were fabricated at the Center for Nano Lithography \& Analysis at The University of Tokyo.
\end{acknowledgments}

\bibliography{Photocurrent}

\end{document}